# Analyzing heterogeneity in Alzheimer Disease using multimodal normative modeling on imaging-based ATN biomarkers

Sayantan Kumar [a,b,*], Tom Earnest [c], Braden Yang [c], Deydeep Kothapalli [c], Andrew J. Aschenbrenner [d], Jason Hassenstab [d], Chengie Xiong [b], Beau Ances [d], John Morris [d], Tammie L. S. Benzinger [c], Brian A. Gordon [c], Philip Payne [a,b], Aristeidis Sotiras [b,c], for the Alzheimer's Disease Neuroimaging Initiative[†]

[a] Department of Computer Science and Engineering, Washington University in St Louis; 1 Brookings Drive, Saint Louis, MO 63130

[b] Institute for Informatics, Data Science & Biostatistics, Washington University School of Medicine in St Louis; 660 S. Euclid Ave, Campus Box 8132, Saint Louis, MO 63110

[c] Mallinckrodt Institute of Radiology, Washington University School of Medicine in St Louis; 4525 Scott Ave, Saint Louis, MO 63110

[d] Department of Neurology, Washington University School of Medicine, 660 S Euclid Ave, Campus Box 8111, St louis, MO 63110

*Corresponding author:

§ sayantan.kumar@wustl.edu, 660 S. Euclid Ave, Campus Box 8132, Saint Louis, MO 63110

[†] Data used in preparation of this article were obtained from the Alzheimer's Disease Neuroimaging Initiative (ADNI) database (adni.loni.usc.edu). As such, the investigators within the ADNI contributed to the design and implementation of ADNI and/or provided data but did not participate in analysis or writing of this report. A complete listing of ADNI investigators can be found at:

http://adni.loni.usc.edu/wpcontent/uploads/how_to_apply/ADNI_Acknowledgement_List.pdf

# Structured Abstract

**INTRODUCTION:** Previous studies have applied normative modeling on a single neuroimaging modality to investigate Alzheimer Disease (AD) heterogeneity. We employed a deep learning-based multimodal normative framework to analyze individual-level variation across ATN (amyloid-tau-neurodegeneration) imaging biomarkers.

**METHODS:** We selected cross-sectional discovery (n = 665) and replication cohorts (n = 430) with available T1-weighted MRI, amyloid and tau PET. Normative modeling estimated individual-level abnormal deviations in amyloid-positive individuals compared to amyloid-negative controls. Regional abnormality patterns were mapped at different clinical group levels to assess intra-group heterogeneity. An individual-level disease severity index (DSI) was calculated using both the spatial extent and magnitude of abnormal deviations across ATN.

**RESULTS:** Greater intra-group heterogeneity in ATN abnormality patterns was observed in more severe clinical stages of AD. Higher DSI was associated with worse cognitive function and increased risk of disease progression.

**DISCUSSION:** Subject-specific abnormality maps across ATN reveal the heterogeneous impact of AD on the brain.

# 1.Background

Alzheimer Disease (AD) is the leading cause of dementia, characterized by cognitive and functional impairments that disrupt daily activities.[1,2] AD is highly heterogeneous, exhibiting considerable variability in clinical manifestations, cognitive decline, disease progression, and neuropathological changes, even within specific diagnostic categories.[3] However, traditional statistical approaches in AD research often overlook this heterogeneity, relying on case-control designs and group averages, effectively treating AD patients as a homogenous group. To progress toward precision medicine in AD, it is essential to move beyond the "average AD patient" approach and the assumption that AD affects all patients in the same way, and characterize disease abnormalities at the individual-level.[4]

Data-driven clustering methods have been the predominant approach for exploring heterogeneity in AD[5]. Normative modeling is an emerging statistical technique that differs from clustering by focusing on subject-level variation instead of group averages.[6–8] Typically, normative analysis in AD research models the relationship between brain measures and covariates (e.g., age, sex) using univariate Bayesian regression models[8,9] or w-scores[10,11], applied to a reference group of healthy controls. The trained normative models are subsequently used to estimate how every AD individual deviates from the norm, resulting in a map of individual-level variability.[8,9] However, normative modeling approaches typically construct separate regression models independently for each brain region, ignoring the multivariate nature of the data. To address this limitation, deep learning approaches based on autoencoders have been used as normative models. These models effectively capture the complex non-linear interactions between brain regions. However, these interactions are typically restricted to same modality measurements.[12–14]

Indeed, most previous studies employing normative models have primarily relied on single modality neuroimaging data to characterize heterogeneity in neuropsychiatric[6,7,15] and neurodegenerative disorders, including AD.[8,16] This is particularly limiting in the case of AD, which is a multi-factorial disease, involving multiple pathological processes that interact and contribute to disease progression. To accurately characterize AD, multi-modal imaging that can quantify biomarker pathology - including amyloid deposition, pathologic tau and neurodegeneration - is essential. Together, these biomarkers compose the National Institute on Aging and Alzheimer's Association research framework that helps with defining AD as a biological construct and facilitates a more comprehensive understanding of individual differences in cognitive performance and clinical progression.[17–21] However, despite recent progress on developing deep learning models for normative modeling across multiple modalities[22–24], these efforts have primarily been methodological and have not investigated AD heterogeneity by taking into account core AD biomarkers (i.e., amyloid and tau).[25]

In this study, we aimed to identify individual patterns of neuroanatomical and neuropathological variation in the brains of individuals with AD using a deep learning based normative modeling framework across amyloid-tau-neurodegeneration (ATN)[26] imaging biomarker data from Alzheimer's Disease Neuroimaging Initiative (ADNI). Accordingly, we trained our previously validated normative modeling framework,[22,24] which is based on multimodal variational autoencoders (mmVAE), on data from a reference control group (i.e., amyloid negative cognitively unimpaired (CU) subjects). We subsequently used the trained model to estimate the extent to which individuals spanning the AD spectrum (ADS) deviate from the normative distribution. Our main objectives can be summarized as follows: (i) assess the extent of neuroanatomical and neuropathological variability between individual patients based on overlapping or distinct patterns of abnormal deviations, (ii) quantify intra-group heterogeneity within ADS clinical groups based on differences in between-participant dissimilarity in abnormal deviations across ATN, (iii) estimate a disease severity index (DSI) for each ADS individual that can capture both the spatial extent of abnormality and the magnitude of regional abnormal deviations across ATN, (iv) examine whether the DSI is related to severity of dementia, impaired cognition and risk of disease progression. The results were replicated in an independent dataset, part of the Charles F. and Joanne Knight Alzheimer's Disease Research Center (ADRC) dataset at Washington University in St. Louis.

## 2. Materials and Methods

### 2.1 Participants

We constructed a discovery dataset consisting of individuals from ADNI and a replication dataset consisting of individuals from the Charles F. and Joanne Knight Alzheimer's Disease Research Center (ADRC) dataset at Washington University in St. Louis. For both datasets, participants were required to have T1-weighted magnetic resonance imaging (MRI), as well as amyloid and tau PET imaging, completed within 1 year of one another. Note that these are cross-sectional cohorts, and we included only the first visit for each individual for which all modalities were available. For both datasets independently, we selected two groups based on amyloid status (Section 2.2.3): (1) a reference control group of amyloid-negative CU (i.e., Clinical Dementia Rating (CDR®) = 0) individuals, which was used to train the deep learning based normative model; and (2) a target disease group of amyloid positive individuals across the ADS. A total of 434 amyloid-negative CU participants were included in the reference control group (ADNI-CU), and 231 amyloid positive individuals across the ADS were included in the target disease group (ADNI-ADS) from ADNI (Figure 1A). For ADRC, the reference group (ADRC-CU) consisted of 301 amyloid negative CU individuals, while the disease group (ADRC-ADS) consisted of 129 amyloid positive individuals on the ADS (Figure 1B).

ADNI-ADS individuals were assigned into 3 diagnostic groups based on CDR: CDR = 0 or preclinical AD (n = 121), CDR = 0.5 (n = 80) and CDR >= 1 (n = 30) (Figure 1A). Similarly, the number of ADRC-ADS individuals in the corresponding groups were 98, 24 and

7 respectively (Figure 1B). The ADNI-CU group was further divided into a training set for model training (ADNI-CU-train; n = 326), a holdout validation set (ADNI-CU-holdout; n = 65) and a test set (ADNI-CU-test; n = 43) at a ratio of 75:15:10 (Figure 1A). The validation set was used to standardize deviations of ADNI-ADS, and calculate Z-scores relative to ADNI-CU. The test sets served as a baseline control group to compare statistics of ADS individuals against amyloid-negative CU participants. Similarly, the ADRC-CU group was also divided into a training set for transfer learning on the replication dataset (ADRC-CU-tl; n = 225), a holdout validation set (ADRC-CU-holdout; n = 45) and a test set (ADRC-CU-test; n = 31) (Figure 1B).

## 2.2 T1-weighted MRI imaging

2.2.1 Image acquisition

ADNI participants included in our analysis underwent T1-weighted MRI imaging using 3T MRI scanners (details are available online https://adni.loni.usc.edu/methods/mri-tool/mri-analysis/). Knight ADRC participants underwent T1-weighted MRI imaging using the Siemens Biograph mMR 3T scanner. Detailed information about MRI image acquisition for the Knight ADRC dataset can be found in the Supplementary Methods (SM1.1).

2.2.2 Image pre-processing

The T1-weighted sequences from ADNI and Knight ADRC were pre-processed using FreeSurfer versions 6 and 5.3, respectively. The cortical surface of each hemisphere was parcellated according to the Desikan–Killiany atlas[27] and anatomical region of interest (ROI) measures were obtained via a whole-brain segmentation procedure (Aseg atlas).[28] The final data included in our analyses (both ADNI and Knight ADRC) included pre-processed regional grey matter volumes of 66 cortical ROIs (33 per hemisphere) and 24 subcortical ROIs for each participant. All ROI volumes were normalized by the intracranial volume (ICV). Detailed MRI pre-processing protocols for ADNI are available online (https://adni.loni.usc.edu/methods/mri-tool/mri-analysis/). Further information about MRI processing protocols for the Knight ADRC dataset can be found in the Supplementary Methods (SM1.2).

## 2.3 Amyloid and tau PET imaging

2.3.1 Image acquisition

ADNI participants underwent amyloid-PET imaging with either [$^{18}$F]-Florbetapir (FBP) or [$^{18}$F]-Florbetaben (FBB) tracers and tau-PET imaging with [$^{18}$F]-Flortaucipir (FTP). Details regarding PET acquisition for ADNI are available online (https://adni.loni.usc.edu/methods/pet-analysis-method/pet-analysis/). Knight ADRC participants underwent amyloid-PET imaging with either FBP or [$^{11}$C]-Pittsburgh Compound B (PIB). Tau-PET imaging was performed using FTP. Further information about PET image acquisition for the Knight ADRC can be found in the Supplementary Methods (SM2.1). To avoid harmonization issues due to multiple tracers, we only included data collected using FBP in our analyses for both ADNI and Knight ADRC.

2.3.2 Image pre-processing

ADNI PET images (FBP and FTP) were registered to the nearest T1-weighted image, which was subsequently processed with FreeSurfer version 6. Detailed PET pre-processing protocols for ADNI can be found available online (https://adni.loni.usc.edu/methods/pet-analysis-method/pet-analysis/). All PET images from Knight ADRC (FBP and FTP) were processed using the PET Unified Pipeline (https://github.com/ysu001/PUP)[29,30]. Further information about PET processing protocols for Knight ADRC can be found in the Supplementary Methods (SM2.2). Similar to MRI ROI volumes, the final amyloid-FBP and tau-FTP data included in our analyses from both ADNI and Knight ADRC consisted of regional standardized uptake value ratio (SUVR) values for 66 cortical ROIs (33 per hemisphere) and 24 subcortical ROIs for each participant. Regional SUVR values in ADNI for amyloid-FBP and tau-FTP were normalized relative to the whole cerebellum and inferior cerebellar gray matter reference region respectively. For Knight ADRC, all regional SUVRs for both amyloid-FBP and tau-FTP were calculated with cerebellum cortex as the reference region.

2.3.3 Amyloid positivity

A summary estimate of global amyloid (FBP) burden in ADNI was calculated as the average SUVR within cortical meta-ROI (spanning frontal, anterior/posterior cingulate, lateral parietal, and lateral temporal regions) which was normalized with a whole cerebellum reference region. ADNI individuals with meta-ROI SUVR uptake greater than 1.11 cut-off were labelled as amyloid positive, following established cut-off procedures recommended within ADNI documentation.[31–34].

For Knight ADRC, an estimate of total cortical amyloid burden was derived by computing the SUVR of a meta-ROI consisting of lateral and medial orbitofrontal, middle and superior temporal, superior frontal, rostral middle frontal, and precuneus ROIs from both hemispheres. The meta-ROI SUVR was normalized using the cerebellar cortex reference region. Knight ADRC individuals with meta-ROI SUVR greater than the 1.24 cut-off were considered amyloid positive, in line with the established literature.[18,35]

## 2.4 Clinical and cognitive assessments

Clinical and cognitive assessments for both ADNI and Knight ADRC were only included if they occurred within one year of MRI imaging. Participants were assessed for dementia using the CDR Scale[36]. Cognitive performance was quantified by computing neuropsychological composites for memory, executive functioning, and language, independently for ADNI and Knight ADRC (SM3).

## 2.5 Multimodal Normative Modeling

2.5.1 Model training

In this study, we used a previously validated deep learning based normative modeling framework.[12,22,24] This framework is based on a multimodal variational autoencoder (mmVAE), which takes as input cross-sectional ATN biomarker data including regional gray matter volumes, as well as amyloid FBP and tau FTP SUVR values (Figure S1). mmVAE had separate encoders for each modality to learn a shared latent space, which is a joint distribution across the different modalities. The shared information in the aggregated latent space was fed through modality-specific decoders to reconstruct each modality (Figure S1). Details of the mmVAE architecture are provided in Supplementary Methods (SM4.1, SM4.2).

Initially, mmVAE was trained using the ADNI-CU-train set, where the input matrix had dimensions of 326 x 270 (subjects x ROIs, with 90 ROIs for each modality). During training, mmVAE learns to reconstruct the multimodal input data as closely as possible to the original. The joint latent distribution allows the model to learn the healthy brain patterns across all modalities. These include the variability in MRI estimates, regional noise or off-target binding for amyloid FBP and age-associated tau FTP accumulation within healthy individuals. mmVAE was conditioned on the age and sex of participants to remove the effect of covariates (see Supplementary Methods SM4.3). For replication in the Knight ADRC dataset, mmVAE pretrained on ADNI-CU-train underwent fine-tuning on ADRC-CU-tl through transfer learning (Section 2.1). Further information about model training and hyperparameter details are available in Supplementary Methods (SM4.3).

2.5.2 Calculating regional deviations for each modality

The main idea of the normative approach is that mmVAE learns only to reconstruct the data of CU individuals. Since individuals on the ADS will differ from CU individuals due to the AD pathology, mmVAE will be less precise in reconstructing their data. As a result, the difference between the reconstructed and input data will be larger in the ADS cohort compared to CU individuals. For each participant in ADNI-ADS and ADRC-ADS, deviations for each region and for each modality can be calculated as the squared error between input and reconstructed data (Figure S1).

2.5.3 Normalizing deviations into Z-scores

Considering the complexity of the brain data, we expect that mmVAE might not fully capture normal variations in healthy subjects, leading to some reconstruction error. Hence, it' is important to standardize ADNI-ADS deviations using ADNI-CU-holdout set (see section 2.1) as a reference for the CU population. We utilized the mean and variance (calculated independently for each region) from ADNI-CU-holdout to normalize regional deviations in the ADNI-ADS cohort across each modality. Following fine-tuning of mmVAE on ADRC-CU-tl, a similar process was applied using ADRC-CU-holdout to normalize deviations in ADRC-ADS. This step generated regional, modality-specific Z-score deviations for each individual in the ADS cohort relative to the normative range of the respective reference control group (Figure S1).

## 2.6 Statistical Analysis

2.6.1 Regional abnormal deviations across ATN biomarkers

For MRI gray matter volumes, regional Z-scores below -1.96 (bottom 2.5% of the normative distribution) were labelled as abnormal (statistically significant) deviations. As followed in previous normative modeling literature[8,16], the lower bound was used because we were interested in gray matter loss (MRI atrophy) associated with neurodegeneration. Similarly regional deviations in amyloid-FBP and tau-FTP SUVR were identified as abnormal if their Z-scores were above 1.96 (top 2.5% of the normative distribution). This upper bound was chosen to focus on increased amyloid/tau SUVR uptake (high amyloid/tau burden) linked to pathological accumulation.[8,16] For each ADS individual, we created a binary thresholded abnormality map, marking regions with abnormal deviations as 1 and others as 0. These abnormality maps were calculated for each modality (MRI, amyloid, or tau) with 90 regions each and also aggregated across all modalities combined (270 regions).

2.6.2 Group differences of regional abnormal deviations across ATN imaging biomarkers

We examined the magnitude of abnormal deviations in each region between amyloid negative CU individuals (CU-test; section 2.1) and clinical groups along the ADS: (i) CDR = 0 (preclinical AD), (ii) CDR = 0.5 (very mild dementia), and (iii) CDR >= 1 (mild or more severe dementia). Our aim was to validate the derived regional abnormal deviations by examine examining whether these deviations showed increased group differences across progressive CDR stages. We quantified group differences using Cohen's d-statistic effect size, calculated separately for each modality.38 A higher effect size when comparing regional MRI volumes indicated lower gray matter volume (more atrophy). Similarly, a higher effect size when comparing amyloid or tau uptake indicated elevated SUVR uptake (higher amyloid and tau loads) compared to the amyloid-negative CU group. We repeated these group comparisons for both ADNI and Knight ADRC independently.

2.6.3 Analysis of spatial distribution of abnormal deviations across ATN across ATN imaging biomarkers

The group comparisons between each of the ADS groups and the amyloid-negative CU group (section 2.6.2) effectively assumed that every clinical group is homogenous in the regional patterns of abnormal deviations across all the modalities. To better understand disease heterogeneity, we also aimed to assess the variability in spatial patterns of neurodegeneration (MRI atrophy) and neuropathology (amyloid and tau deposition) across clinical groups. Towards this end, we computed the proportion of abnormal deviations (fraction of individuals with abnormal deviations; section 2.6.1) separately for each region, modality and clinical group. The regional proportion of abnormal deviations was calculated for each of the ADS clinical groups and a group of amyloid-negative CU individuals (CU-test; section 2.1) independently for both the ADNI and Knight ADRC datasets.

2.6.4 Intra-group heterogeneity within ADS groups

Next, we aimed to quantify the intra-group heterogeneity across ADS clinical groups and assess whether this increases across progressive dementia stages. We used hamming distance to measure the dissimilarity in binary-thresholded abnormality maps (section 2.6.1) between every pair of ADS individuals, with higher distance indicating more dissimilarity. Hamming distances were estimated for both modality-specific abnormality maps (hamming_mri, hamming_amyloid and hamming_tau) and abnormality maps aggregated across all modalities (hamming_all). Median hamming distances were compared across ADS groups and an amyloid-negative CU group (CU-test; section 2.1). Distribution of hamming distances were visualized using kernel density estimation (KDE) plots across ADS groups, reflecting the extent of intra-group heterogeneity. The intra-group heterogeneity analysis was performed independently for both ADNI and Knight ADRC cohorts.

2.6.5 Disease severity index (DSI) across ADS groups

Our objective was to design a Disease Severity Index (DSI) for each ADS individual which can capture both the spatial extent and magnitude of regional abnormal deviations across multiple modalities into a single, subject-specific metric. DSI was calculated separately for each modality (DSI_mri, DSI_amyloid, DSI_tau) and also aggregated across all modalities (DSI_all). Specifically, every individual's DSI was calculated by i) first performing an inner product between the binary thresholded abnormality map (section 2.6.1) and the regional deviation vector; and then normalizing the inner product by the total number of regions ($n_R = 90$ for DSI_mri, DSI_amyloid, DSI_tau and $n_R = 270$ for DSI_all).

DSI_all represents a personalized measure of brain health that accounts for individual variability in gray matter volume, and amyloid and tau deposition, rather than relying on average group relationships. To demonstrate this, we first compared DSI values between the different ADS groups and a group of amyloid-negative CU individuals (CU-test; section 2.1). This allowed us to examine the association between increasing DSI and progressive dementia stages (high CDR). FDR-corrected post hoc Tukey comparisons were used for pairwise group differences.

2.6.6 Association between DSI and cognitive performance

We then examined the associations between DSI values (DSI_mri, DSI_amyloid, DSI_tau and DSI_all) and the three neuropsychological composites: memory, executive functioning, and language (section 2.4, SM3) in both ADNI-ADS and ADRC-ADS cohorts. The associations were estimated using linear regression, adjusted for age and sex. Additionally, Pearson correlation coefficient was used to measure the pairwise correlation between each of the DSI categories and the composites.

2.6.7 Relationship between DSI and CDR progression

Lastly, we examined associations of DSI_all and clinical progression in both ADNI-ADS and ADRC-ADS cohorts. For this analysis, we included subjects with follow-up CDR status data, who were CDR $< 1$ at their baseline visit with all three modalities present. We analyzed the relationship between DSI_all and CDR progression using survival analysis, adjusting for age and sex. The event of interest was progression to CDR $>= 1$. A Kaplan-Meier plot was used to illustrate the impact of the 4 DSI_all quantiles on disease progression risk. Log-rank tests estimated pairwise differences in progression risk among the DSI_all quantiles. Post-hoc comparisons were adjusted for multiple comparisons using FDR.[37]

<u>2.6.8 Code availability and visualizations</u>

All analyses were performed using Python 3.7. All visualizations of the brain atlases for the effect size maps (Section 2.5.2) and proportion of abnormal deviation maps (Section 2.5.3) were visualized using the python package ggseg.[38] Hamming distance distributions at group level were visualized using kernel density estimation (KDE) plots. Code for the project will be made publicly available upon acceptance.

# 3. Results

## 3.1 Dataset characteristics

Sample characteristics for the ADNI-ADS and ADRC-ADS cohorts are shown in Table 1. The ADNI-ADS cohort was older ($p = 0.035$), while the ADRC-ADS cohort had more females ($p = 0.006$). ADRC-ADS showed less memory impairment, indicated by higher MMSE scores ($p < 0.001$) and a higher proportion of individuals with CDR = 0 or preclinical AD ($p < 0.001$). Sample characteristics for the ADNI-CU and ADRC-CU datasets are in Table S1. ADNI-CU participants were older ($p < 0.001$), whereas there were more females in ADRC-CU ($p = 0.026$). ADNI-CU participants had slightly lower MMSE scores than ADRC-CU, but this difference was not statistically significant ($p = 0.067$). CU participants in ADNI and ADRC had lower age and higher MMSE scores compared to their ADS counterparts (i.e., ADNI-CU vs. ADNI-ADS; ADRC-CU vs. ADRC-ADS), with all differences being statistically significant ($p < 0.001$).

**Table 1**: Descriptive statistics for the ADNI-ADS and ADRC-ADS datasets. Statistical differences were assessed using two-sided ANOVA (continuous variables) and chi-squared tests (categorical. variables). Significant p-values are highlighted in bold with *: $0.01 < p < 0.05$, **: $0.005 < p < 0.01$, ***: $p < 0.001$. Abbreviations: SD = standard deviation, ANOVA = analysis of variance, CDR = Clinical Dementia Rating, MMSE = Mini-Mental State Examination.

|  | **ADNI-ADS** | **ADRC-ADS** | **p-value** |
|---|---|---|---|
| **N** | 231 | 129 | - |
| **Sex, Male: Female** | 108:123 | 48:81 | **p = 0.035*** |
| **Age (mean +/- SD)** | 73.6 +/- 6.9 | 71.5 +/- 8.3 | **p = 0.006**** |
| **CDR (0/0.5/>=1)** | 121/80/30 | 98/24/7 | **p < 0.001**** |
| **MMSE (mean +/- SD)** | 24.5 +/- 3.2 | 26.5 +/ 3.7 | **p < 0.001 **** |

## 3.2 More severe dementia was associated with pronounced regional atrophy and elevated regional amyloid and tau burden in ADS patients

3.2.1 Discovery dataset - ADNI

Region-level (total of 90 regions—FDR corrected) pairwise group comparisons with amyloid-negative CU individuals (CU-test) provided evidence that gray matter volumes were lower in 56 regions in mild or more severe dementia, in 22 regions in very mild dementia, and no regions in preclinical AD (Table S2). Maximum group differences in atrophy were observed in the temporal, parietal, and hippocampal regions, and to a lesser extent in the frontal, occipital, and amygdala regions (Figure 2A). Expectedly, regional-level pairwise group comparisons with amyloid-negative CU individuals revealed higher amyloid burden in 84 regions in mild or severe dementia, in 75 regions in very mild dementia, and 85 regions in preclinical AD (Table S2). Higher effect sizes for amyloid burden were mostly observed in the medial orbitofrontal, precuneus, temporal and frontal pole regions. Similarly, pairwise group comparisons with CU individuals revealed increased tau deposition in 80 regions in mild or severe dementia, in 62 regions in very mild dementia, and no regions in preclinical AD (Table S2). Regions with high effect sizes for tau burden included the temporal, frontal, precuneus, parietal and hippocampal regions (Figure 2A).

3.2.2 Replication dataset - Knight ADRC

We found a notable similarity between the effect size maps estimated in ADRC-ADS and in ADNI-ADS (Figure 2B). Similar to ADNI-ADS, statistically significant volumetric differences in ADRC-ADS were mainly seen in the temporal, parietal, and hippocampal regions. More regions showed pronounced atrophy in mild or severe dementia ($n_r = 50$) compared to very mild dementia ($n_r = 25$), with no abnormal regions for preclinical AD. Elevated amyloid burden was observed in more regions for mild or severe dementia ($n_r = 81$) compared to very mild dementia ($n_r = 77$), with 82 abnormal regions for preclinical AD (Table S2). Greater effect sizes were observed in the frontal and temporal regions (Figure 2B). Similarly, statistically significant group differences in regional tau deposition were found across various dementia severity levels (mild or severe dementia; $n_r = 74$, very mild dementia; $n_r = 52$, and preclinical AD; $n_r = 0$), with greater effect sizes observed in the temporal and hippocampal regions (Table S2, Figure 2B).

## 3.3 ADS individuals with more severe dementia have higher proportion of abnormal deviations in regional atrophy, amyloid and tau burden

3.3.1 Discovery dataset - ADNI

The proportion of abnormal deviations defined within each clinical group differed in regional patterns between the mild or more severe dementia, the very mild dementia, the preclinical AD, and the controls group (Figure 3A). As far as regional gray matter volumes are concerned, the highest proportion of abnormal deviations was observed in hippocampal regions: 47% in the mild or more severe dementia group, 25% in the very mild dementia group, 6% in the preclinical AD group, and 3% in the amyloid-negative CU group. Regarding regional amyloid burden, the highest proportion of abnormal deviations was observed for precuneus and frontal pole cortices: 100% in the mild or more severe dementia group, 87% in the very mild dementia group, 71% in the preclinical AD group, and 5% in the CU group. Lastly, the proportion of abnormal deviations in tau deposition was observed in hippocampal and entorhinal regions: 84% in the mild or more severe to severe dementia group, 65% in the very mild dementia group, 24% in the preclinical AD group, and 14% in the CU group. Overall, a higher proportion of abnormal deviations was observed for regional amyloid and tau burden than gray matter volume. This trend was consistent across the dementia stages (Figure 3A).

3.3.2 Replication dataset - Knight ADRC

In line with the ADNI-ADS results, we observed a consistent pattern of higher proportion of regional abnormal deviations with increased dementia severity (Figure 3B). Additionally, we observed similar regional patterns of abnormal deviations across all three modalities. Specifically, the highest proportion of abnormal deviations was observed for the same regions in both datasets, i.e., hippocampal and temporal regions for MRI, frontal regions and precuneus for amyloid, and temporal and parietal regions for tau (Figure 3B) However, a lower proportion or regional abnormal deviations was observed for the mild or more severe dementia group in ADRC-ADS compared to the corresponding group in ADNI-ADS. This is likely due to the smaller sample size for this group (n = 7) in ADRC-ADS (Figure 3B).

### 3.4 ADS individuals with more severe dementia are more heterogenous compared to individuals with less dementia

3.4.1 Discovery dataset - ADNI

The distribution of Hamming distance calculated for all modalities (hamming_all) showed greater within-group heterogeneity (dissimilarity) for ADNI-ADS individuals at progressive stages of dementia (Figure 4A). The median Hamming distance (hamming_all) significantly differed between groups overall (p < 0.001). Pairwise comparisons in median hamming distance (Tukey post-hoc) were all significant (p < 0.001) except between the very mild dementia and the mild or more severe dementia group. Specifically, the Hamming distance was the highest in individuals with mild or more severe dementia (median 62, IQR 39, 95% CI 60.3–63.8), followed by the very mild dementia group (median 56, IQR 34, 95% CI 58.1–59.3) and the preclinical AD (CDR = 0) group (median 47, IQR 30, 95% CI 49.2–50.1) (Figure 4A). The lowest Hamming distance was observed in the CU group (ADNI-CU-test; median 5, IQR 6, 95% CI 8.4–10.7). For Hamming distance variants calculated using a single modality, we observed the same pattern of higher dissimilarity for ADS patients at progressive stages of dementia (Figure S2A). Within each clinical group, MRI showed the highest within-group heterogeneity in spatial patterns of abnormal deviations, followed by tau and amyloid (Figure S2A).

3.4.2 Replication dataset - Knight ADRC

Consistent with ADNI-ADS results, we observed greater within-group dissimilarity for ADRC-ADS individuals at progressive stages of dementia (Figure 4B). Similar to ADNI, the median Hamming distance (hamming_all) significantly differed between groups overall (p < 0.001). Pairwise comparisons in median hamming distance (Tukey post-hoc) were all significant (p < 0.001) except between the very mild dementia and the mild or more severe dementia group. Within-group dissimilarity was higher for the mild or more severe (median 39, IQR 40, 95% CI 28.8–43.5) and the very mild dementia (median 30, IQR 33, 95% CI 30.7–34.5) groups compared to preclinical AD (median 14, IQR 20, 95% CI 21.6–22.9) and the CU groups (median 4, IQR 8, 95% CI 6.9–8.5). We observed the same trend when examining Hamming distances calculated for each modality separately (hamming_mri, hamming_amyloid, and hamming_tau). Lastly, similarly to ADNI-ADS results, MRI exhibited the highest within-group dissimilarity in spatial patterns of abnormal deviations compared to amyloid and tau (Figure S2B).

**3.5 Higher DSI was associated with progressive stages of dementia**

3.5.1 Discovery dataset - ADNI

DSI calculated across all modalities (DSI_all) exhibited minimal values for CU individuals (ADNI-CU-test; mean = 0.06, IQR = 0.03, 95% CI = [0.007-0.1]), with a consistently increasing trend across dementia stages (Figure 5A). Specifically, maximum DSI_all values were observed for ADNI-ADS individuals with mild or more severe dementia (mean = 1.8, IQR = 1.5, 95% CI = [1.31-2.26]), followed by individuals with very mild dementia (mean = 1.1, IQR = 1.3, 95% CI = [0.87-1.25]) and preclinical AD individuals (mean = 0.45, IQR = 0.6, 95% CI = [0.39-0.52]). Pairwise group differences were statistically significant (FDR corrected $p < 0.05$), except between very mild dementia and mild to more severe dementia groups (Figure 5A). When calculated across individual modalities, DSI values consistently increased at more advanced stages of dementia for all three modalities (i.e., DSI_mri, DSI_amyloid, DSI_tau; Figure S3A). DSI values calculated for amyloid and tau (DSI_amyloid, DSI_tau) were higher across CDR groups compared to DSI values calculated for MRI (DSI_mri). Pairwise group differences were statistically significant for DSI calculated for individual modalities except between very mild dementia and moderate to severe dementia groups for DSI_mri, DSI_amyloid, DSI_tau, and between CU-test and preclinical AD for DSI_mri and DSI_tau (Figure S3A).

3.5.2 Replication dataset – Knight ADRC

We found similar patterns of increasing DSI values at progressive dementia stages in the ADRC-ADS cohort (Figure 5B). DSI_all was highest for individuals with mild to more severe dementia (mean = 0.46, IQR = 0.28, 95% CI = [0.25-0.6]) and lowest for ADRC-CU-test individuals (mean = 0.006, IQR = 0.005, 95% CI = [0.002-0.008]) (Figure 5B). Individuals with preclinical AD (mean = 0.21, IQR = 0.22, 95% CI = [0.12-0.37]) and very mild dementia exhibited intermediate DSI_all values (mean = 0.32, IQR = 0.32, 95% CI = [0.21-0.5]). Pairwise group differences were statistically significant ($p < 0.05$) except between the very mild dementia and moderate to severe dementia groups (Figure 5B). Notably, higher DSI_all values were observed in ADNI-ADS compared to ADRC-ADS, likely due to more ADS individuals with advanced disease stages in ADNI compared to the ADRC dataset. Modality-specific DSI values (i.e., DSI_mri, DSI_amyloid, and DSI_tau) for ADRC-ADS individuals showed similar trends as observed in ADNI (Figure S3B).

### 3.6 Higher DSI values were associated with impaired cognition

3.6.1 Discovery dataset - ADNI

Linear regression, adjusted for age and sex, revealed significant associations between higher DSI_all values and decreased values in neuropsychological composites: memory ($\beta = -0.6$; $p < 0.001$; $r = -0.62$), executive functioning ($\beta = -0.46$; $p < 0.001$; $r = -0.54$), and language ($\beta = -0.39$; $p < 0.001$; $r = -0.47$) (Table 2). Similar statistically significant associations were observed for DSI values based on individual modalities (DSI_mri, DSI_amyloid, DSI_tau). The correlations were higher for DSI_tau compared to DSI_mri and DSI_amyloid (Table S3). However, correlations were higher for DSI_all compared to modality-specific DSI, highlighting the benefits of taking into account information across all modalities (Table 2).

3.6.2 Replication dataset - Knight ADRC

We observed similar trends of significant associations between DSI_all and the neuropsychological composites in ADRC-ADS (Table 2). In ADRC-ADS, DSI_all was significantly associated with memory ($\beta = -0.71$; $p < 0.001$; $r = -0.68$), executive functioning ($\beta = -0.52$; $p < 0.001$; $r = -0.56$), and language ($\beta = -0.36$; $p < 0.001$; $r = -0.41$). Similarly, DSI calculated for individual modalities exhibited significant associations with cognitive domains, with DSI_amyloid and DSI_tau generally showing higher correlations than DSI_mri (Table S3).

**Table 2**: Comparison between DSI across all modalities (DSI_all), and the ATN summary metrics (hippocampal volumes, amyloid burden, and tau index) with respect to association with the composite cognitive scores (memory, executive functioning, and language). β represents the slope and p represents the p-value for linear regression, adjusted for age and sex. r represents the Pearson correlation coefficient.

|  | Cognitive domain | ADNI-ADS | | | ADRC-ADS | | |
| --- | --- | --- | --- | --- | --- | --- | --- |
|  |  | β | p | r | β | p | r |
| **DSI_all** | **Memory** | - 0.65 | p < 0.001 | **- 0.62** | - 0.71 | p < 0.001 | **- 0.68** |
|  | **Executive** | - 0.46 | p < 0.001 | **- 0.54** | - 0.52 | p < 0.001 | **- 0.56** |
|  | **Language** | - 0.39 | p < 0.001 | **- 0.47** | - 0.36 | p < 0.001 | **- 0.41** |

### 3.7 Higher DSI is associated with increased risk for clinical progression

3.7.1 Discovery dataset - ADNI

Longitudinal clinical status data were available for 175 individuals in ADNI-ADS with either no or very mild dementia at baseline. DSI_all was significantly associated with the risk of progressing to mild or more severe dementia (p < 0.001; Figure 6A). Notably, individuals in higher DSI quartiles, particularly q4 (p < 0.001) and q3 (p < 0.01), demonstrated a heightened risk of progression compared to those in lower quartiles, namely q1 and q2 (Table S4).

3.7.2 Replication dataset - Knight ADRC

Longitudinal clinical status data were available for 85 ADS individuals with either no or very mild dementia at baseline (CDR = 0 or CDR = 0.5). As in ADNI-ADS, the survival analysis in ADRC-ADS showed an association between DSI_all and clinical progression (Figure 6B). In ADRC-ADS, individuals in q4 (p < 0.001) and q3 (p < 0.01) progressed more rapidly to severe dementia than those in q1 and q2 (Table S4).

# 4. Discussion

In this study, we applied a deep learning based normative modeling framework across multiple neuroimaging modalities to assess heterogeneity in neuroanatomical and neuropathological changes in the brain of individuals with AD. Results showed evidence of (i) heterogeneous patterns of abnormal deviations in regional volumetric measurements as well as amyloid and tau deposition between patients with AD; (ii) increased dissimilarity in spatial patterns of abnormal deviations for AD patients at more severe dementia stages; (iii) associations of DSI, which distils spatial patterns of abnormal deviations across multiple modalities in a single index for each subject, with cognitive performance, as well as (iv) associations of DSI with increased risk of disease progression. Our observations were reproducible in both the discovery and replication datasets, which demonstrated the generalizability of our scientific findings.

## 4.1 Deep learning based normative modeling

Normative approaches to study heterogeneity in AD typically learn a regression model independently for each brain region, ignoring the multivariate nature of the data.[4,8,16] Our work used a deep learning-based normative model, specifically a variational autoencoder, to capture the complex non-linear interactions within multivariate data, rather than modeling each variable independently. However, existing studies using deep normative models have been limited in studying interactions between single modality measurements (e.g., regional volumetric measurements extracted from T1-weighted MRI).[12,13] To accurately characterize a multi-factorial disease like AD, it is essential to use multiple neuroimaging modalities that can quantify biomarker pathology - including neurodegeneration, amyloid and tau deposition. Despite recent progress on developing deep learning normative models for multiple modalities, these approaches have primarily focused on methodological advancements.[14,23] Further, these studies validated their approach using measurements extracted from MRI and did not investigate AD heterogeneity using the core AD biomarkers (i.e. amyloid and tau). Our work used a multimodal deep learning based normative modeling framework to investigate AD heterogeneity through the lens of multimodal imaging-based ATN biomarkers i.e., neurodegeneration, amyloid and tau biomarkers.

## 4.2 Variation in spatial patterns of abnormal deviations across gray matter volume, amyloid and tau burden

Our findings both complement and provide new insights to the established understanding of the neurobiology of AD. High proportion of abnormal deviations in regional gray matter volumetric measurements was observed in hippocampal and medial-temporal regions, areas known to be associated with neurodegeneration.[39,40] Abnormal deviations in amyloid deposition were observed the most in the in the precuneus, frontal and temporal regions, which are among the first areas to accumulate FBP amyloid pathology.[41–43] Similarly, maximum abnormal deviations in tau deposition were observed in the medial and lateral temporal regions. These regions are among the first to accumulate tau and are typically used to construct the meta-ROI to characterize early tau accumulation.[10,44]

The regions with the highest proportion of abnormal deviations across all modalities (e.g., medial-temporal regions) represent the areas associated with onset of clinical symptoms related to "typical AD". The proportion of abnormal deviations for these "typical" regions also increases with disease severity which complements the current literature on AD.[39,40] However, our work extends the literature by showing that there are other regions with lower proportion of abnormal deviations. This suggests partial overlap between ADS patients, which challenges the validity of a "typical AD" patient. Further, this is in conflict with the assumption of disease group homogeneity that is common among typical analytical tools, such as case-control studies. Further evidence for the partial overlap between patients were provided by the Hamming distance analyses, which quantified it and demonstrated increased dissimilarity in spatial patterns of abnormal deviations between AD patients at more severe dementia stages. Together, the results of the regional proportion of abnormal deviations across the three modalities (Figure 3) and the results of the Hamming distance analyses (Figure 4) provide evidence that ADS patients not only differ in the number of regions with statistically significant abnormal deviations but also in their respective patterns of abnormal deviations.

Further, our results indicated that for all ADS groups, the spatial patterns of abnormal deviations for MRI had the highest within-group dissimilarity, followed by tau and amyloid (Figure S2). These observations are also supported by our results in Figure 3 where the highest proportion of abnormal deviations across all ADS groups was the lowest for MRI (47%), higher for tau (84%) and the highest for amyloid (100%). This indicated greater variation in the spatial patterns of gray matter atrophy and tau pathology compared to amyloid deposition.

The observed variation in abnormal deviations across the three imaging modalities is in line with previous single-modality normative modeling and subtyping studies.[16,45,46] Specifically, a similar proportion of regional abnormal deviations was reported in a previous normative modeling study examining cortical thickness heterogeneity in AD.[16] Moreover, studies using T1-weighted MRI or tau PET to estimate subtypes within the ADS population identified more subtypes for MRI (4 subtypes[47–49] or 3 subtypes[50–53]) and tau (4 subtypes[45,54,55]) compared to the ones using amyloid PET as the input modality (2 subtypes[46,56]). The fewer amyloid-driven subtypes indicate less heterogeneity compared to MRI and tau, consistent with our findings.

## 4.3 DSI_all as a potential marker of brain health

We developed a multimodal metric DSI_all which provides an individualized metric of brain health that takes into account individual variability in patterns of gray matter volume, amyloid and tau deposition instead of quantifying group average relationships. Recent studies have relied on single modality to calculate the total count of regions with abnormal deviations in cortical thickness as a marker of disease progression.[8,16] Additionally, recent studies have also quantified tau spread (TSS) for predicting cognitive impairment and disease progression.[57,58] In contrast to these previous approaches, DSI_all captures both the spatial spread and magnitude of regional abnormal deviations across all modalities. This allows DSI_all to quantify neurodegenerative and neuropathological changes in the brain, providing a personalized metric of brain health, which can assist in clinical decision making. This is further supported by the demonstrated associations with cognitive performance, disease severity and clinical progression. Lastly, DSI_all can be potentially used to monitor the amyloid burden and track patient response to recently FDA-approved AD treatments, such as Aducanumab and Lecanemab medications.[59,60]

### 4.4 Limitations, and scope for future work

There are certain limitations that need to be considered regarding our analyses. First, we used cross-sectional imaging data for our normative modeling framework, providing a snapshot of the disease at a specific time. Due to the cross-sectional setting, it is challenging to distinguish between stages and subtypes. While the results of our analyses examining the regional proportions of abnormal deviations as well as patient dissimilarity across disease stages demonstrated that the observed variability of spatial patterns is due to both disease progression and spatial heterogeneity, future works should incorporate serial neuroimaging data collected across multiple time points to better characterize disease progression and heterogeneity. Second, both ADNI and Knight ADRC datasets consisted of individuals from North America only, which are not representative of the general population. Future normative modeling works should consider a more diverse population for the reference control group, sampled from different geographical regions. This can allow studies to have a larger sample size for a more accurate representation of the healthy brain. Third, the imaging scans in the discovery and replication cohorts were processed with a different version of FreeSurfer (FreeSurfer 6 for

ADNI and FreeSurfer 5.3 for ADRC). Although we fine-tuned our pre-trained normative model on the replication cohort, the different versions of FreeSurfer may potentially add noise to the normative model. To address this issue, it is important to consider harmonization methods like COMBAT for future multi-site studies on normative modeling.[61]

## 4.5 Conclusions

In this paper, we assessed the heterogeneity in AD through the lens of multiple neuroimaging modalities by estimating regional statistically significant neurodegenerative and neuropathological deviations at the individual level. We studied these subject-specific maps of regional abnormal deviations across gray matter volume, amyloid burden and tau deposition and observed higher variability in the spatial patterns of MRI atrophy compared to amyloid and tau burden. Additionally, we showed higher within-group heterogeneity for ADS patients at increased dementia stages. Lastly, we developed an individualized metric of brain health that summarizes the extent and severity of neurodegeneration and neuropathology. Together the individualized disease severity index and the subject-specific maps of abnormal deviations have the potential to assist in clinical decision making and monitor patient response to anti-amyloid treatments. Our results were reproducible in both the discovery and replication datasets, demonstrating the generalizability of our findings.

# Acknowledgements

We would like to thank the staff for the Washington University Center for High Performance Computing who helped enable this work.

# Competing Interests

Author AS has received personal compensation for serving as a grant reviewer for BrightFocus Foundation. The remaining authors have no conflicting interests to report.

# Funding Sources

The preparation of this report was supported by the Centene Corporation contract (P19-00559) for the Washington University-Centene ARCH Personalized Medicine Initiative and the National Institutes of Health (NIH) (R01-AG067103). Computations were performed using the facilities of the Washington University Research Computing and Informatics Facility, which were partially funded by NIH grants S10OD025200, 1S10RR022984-01A1 and 1S10OD018091-01. Additional support is provided by the McDonnell Center for Systems Neuroscience.

Data collection and sharing for this project was funded by the Alzheimer's Disease Neuroimaging Initiative (ADNI) (National Institutes of Health Grant U01 AG024904) and DOD ADNI (Department of Defense award number W81XWH-12-2-0012). ADNI is funded by the National Institute on Aging, the National Institute of Biomedical Imaging and Bioengineering, and through generous contributions from the following: AbbVie, Alzheimer's Association; Alzheimer's Drug Discovery Foundation; Araclon Biotech; BioClinica, Inc.; Biogen; Bristol-Myers Squibb Company; CereSpir, Inc.; Cogstate; Eisai Inc.; Elan Pharmaceuticals, Inc.; Eli Lilly and Company; EuroImmun; F. Hoffmann-La Roche Ltd and its affiliated company Genentech, Inc.; Fujirebio; GE Healthcare; IXICO Ltd.; Janssen Alzheimer Immunotherapy Research & Development, LLC.; Johnson & Johnson Pharmaceutical Research & Development LLC.; Lumosity; Lundbeck; Merck & Co., Inc.; Meso Scale Diagnostics, LLC.; NeuroRx Research; Neurotrack Technologies; Novartis Pharmaceuticals Corporation; Pfizer Inc.; Piramal Imaging; Servier; Takeda Pharmaceutical Company; and Transition Therapeutics. The Canadian Institutes of Health Research is providing funds to support ADNI clinical sites in Canada. Private sector contributions are facilitated by the Foundation for the National Institutes of Health (www.fnih.org). The grantee organization is the Northern California Institute for Research and Education, and the study is coordinated by the Alzheimer's Therapeutic Research Institute at the University of Southern California. ADNI data are disseminated by the Laboratory for Neuro Imaging at the University of Southern California.

Data were also provided (in part) by Knight Alzheimer Disease Research Center (ADRC), supported by the Alzheimer's Disease Research Center grant [P50-AG05681], Healthy Aging and Senile Dementia [P01 AG03991], and Adult Children Study [P01

## Author contributions

All authors contributed to the conceptualization and design of the study. SK implemented all data analyses and experiments and wrote the first draft of the manuscript. AS contributed to the interpretation of data. TE, BG and DK provided technical support. All authors were involved with manuscript revision, and all approved of the final draft.

## Data availability and consent statement

All ADNI participants provided written informed consent, and study protocols were approved by each local site's institutional review board. ADNI data used in this study are publicly available and can be requested following ADNI Data Sharing and Publications Committee guidelines: https://adni.loni.usc.edu/data-samples/access-data/. All protocols for Knight ADRC were approved by the Institutional Review Board at Washington University in St. Louis, and all participants provided informed consent before all procedures. Knight ADRC data can be obtained by submitting a data request through https://knightadrc.wustl.edu/data-request-form/.

## Figures

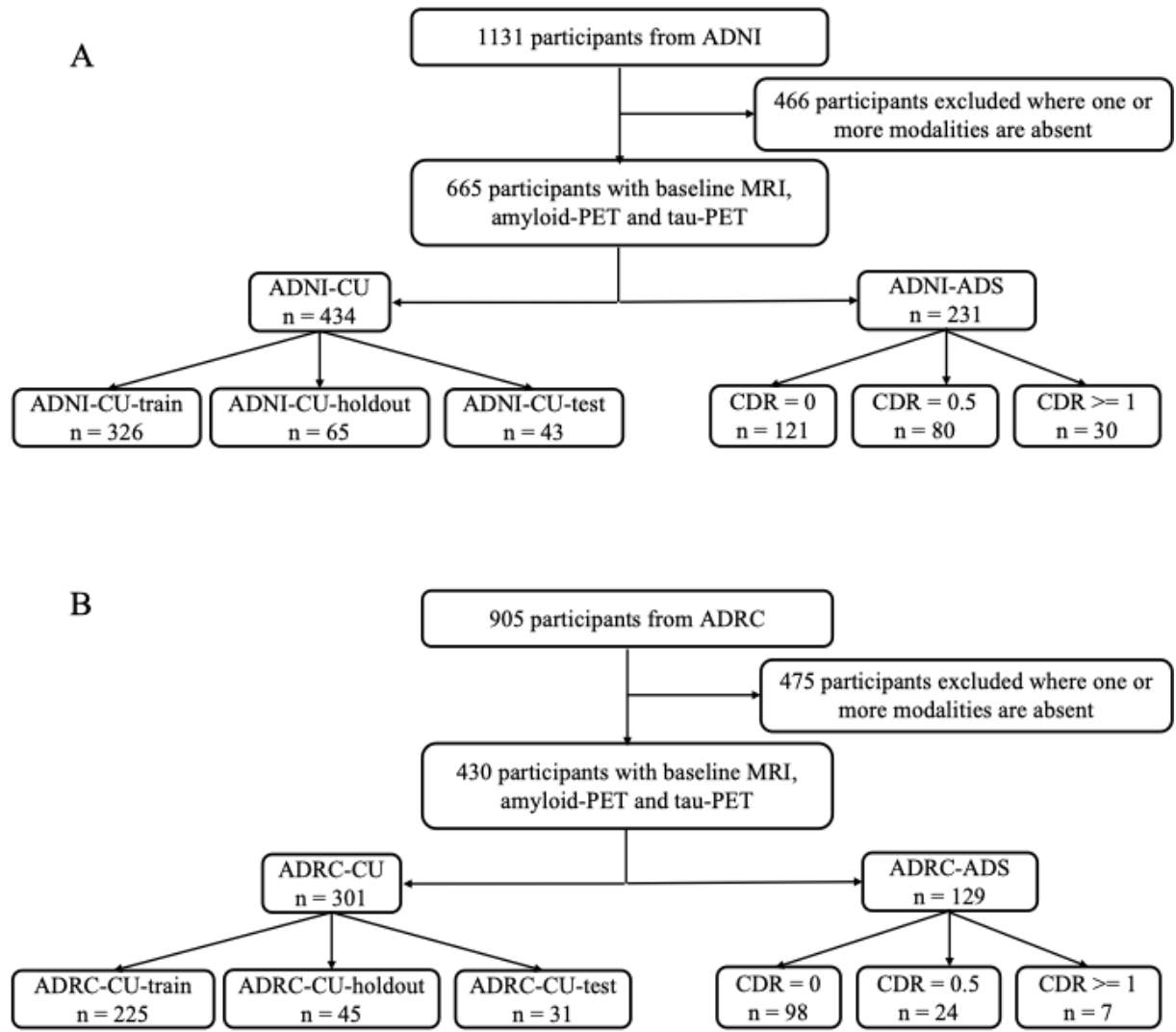

**Figure 1**: Flow chart of ADNI (**1A**) and Knight ADRC (**1B**) study participants.

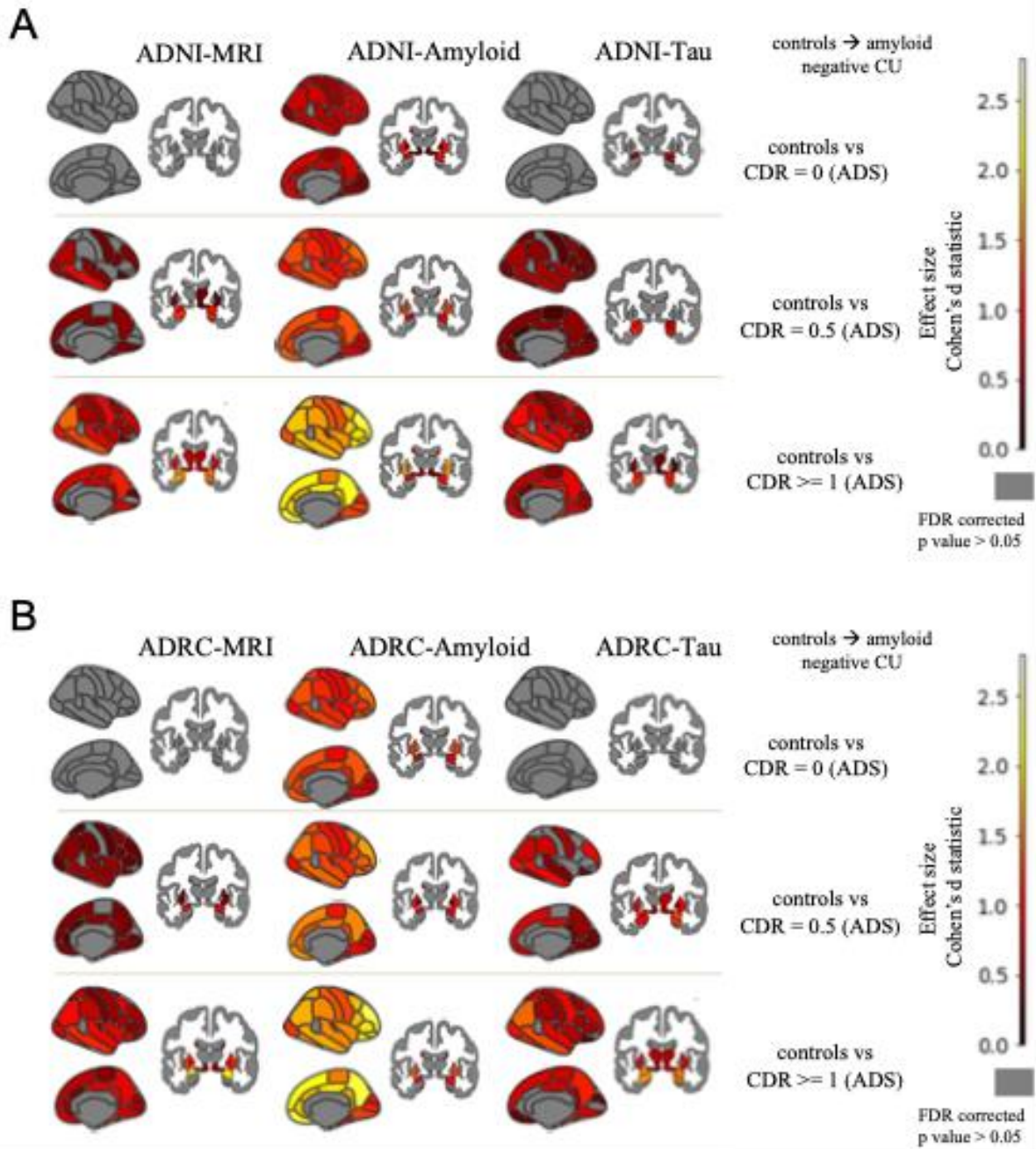

**Figure 2**: Brain atlas maps (Desikan-Killiany atlas for 66 cortical regions and Aseg atlas for 24 subcortical regions) showing the pairwise group differences in magnitude of deviations at each region between the amyloid negative CU group and each of the CDR groups in ADNI (**2A**) and Knight ADRC (**2B**). The figures from left to right indicate the brain maps corresponding to MRI, amyloid and tau, respectively. The color bar represents the effect size (Cohen's d statistic). Effect sizes of d = 0.2, d = 0.5, and d = 0.8 are typically categorized as small, medium, and large, respectively. Gray regions represent the regions with no statistically significant deviations after FDR correction.

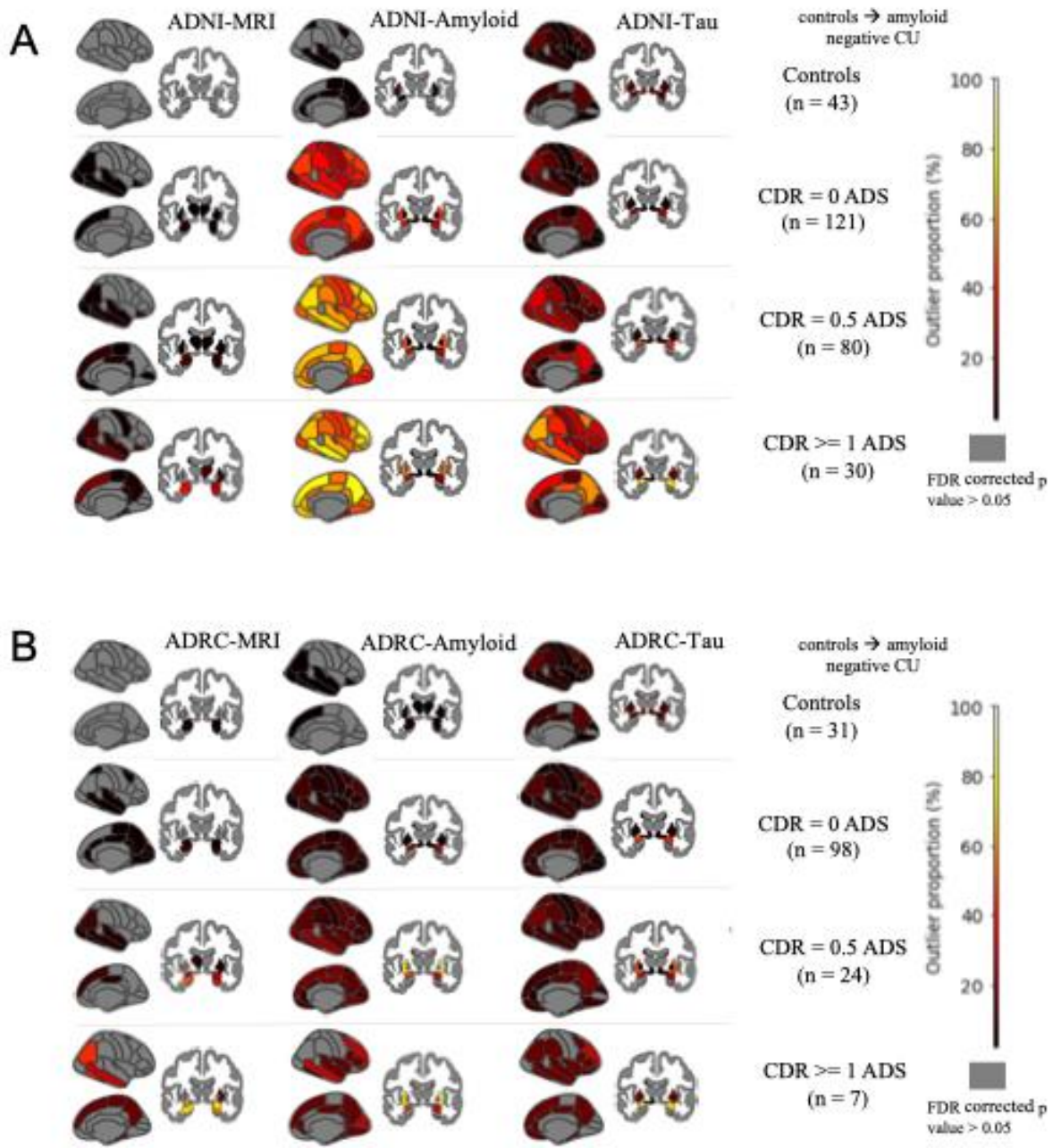

**Figure 3**: Brain atlas maps (Desikan-Killiany atlas for 66 cortical regions and Aseg atlas for 24 subcortical regions) showing the proportion of abnormal deviations for each region in ADNI (**3A**) and Knight ADRC (**3B**). The figures from left to right indicate the brain maps corresponding to MRI, amyloid and tau respectively. The color bar represents the proportion of abnormal deviations of each region from 0 to 100%. Gray represents that no participants have abnormal deviations for that region.

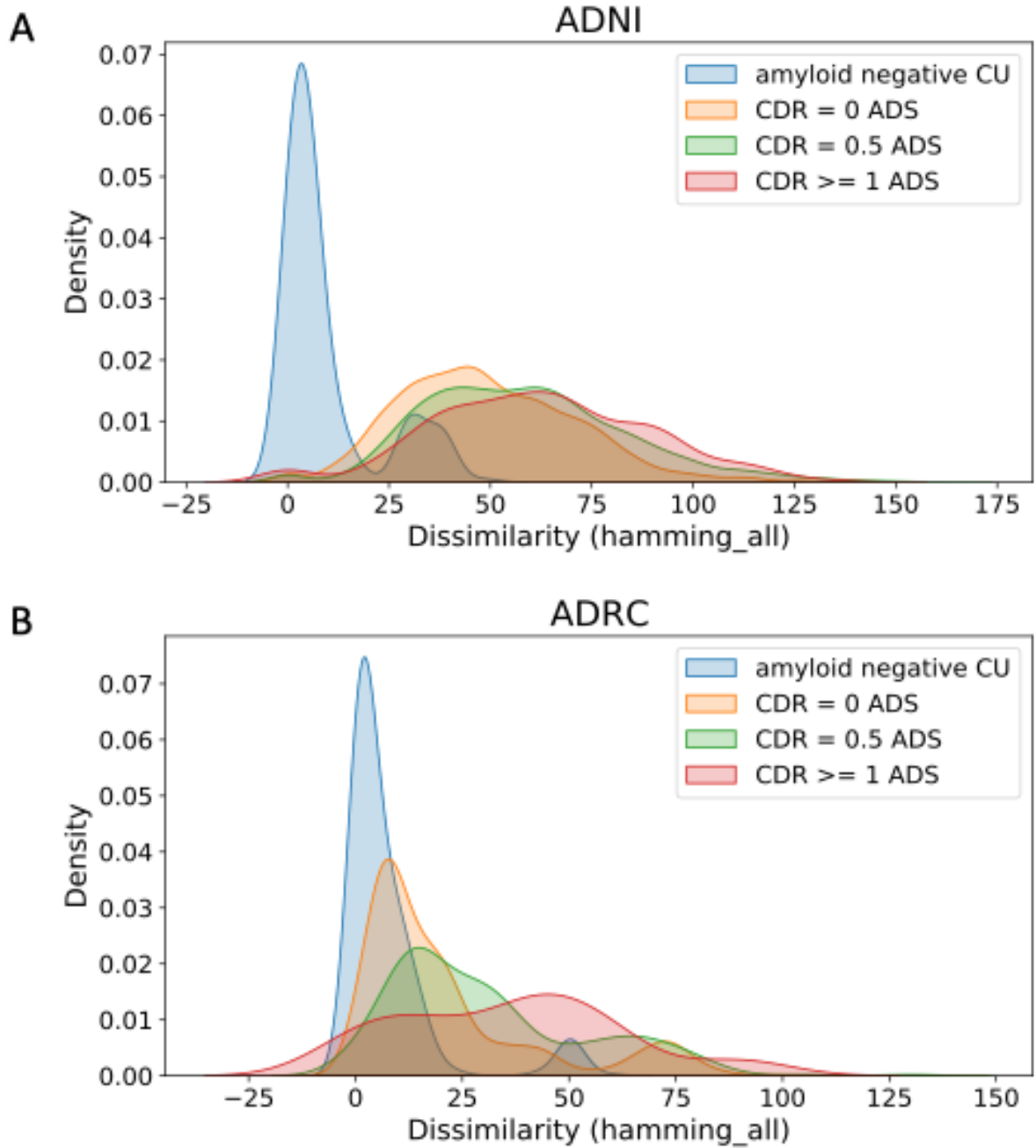

**Figure 4**: Hamming distance density (KDE plot) which illustrates the spread of dissimilarity in abnormality patterns (calculated by the Hamming distance for all modalities or hamming_all; see Section 2.5.4) within each CDR group for ADNI (**4A**) and Knight ADRC (**4B**). Higher hamming distance values indicated intra-group more heterogeneity in abnormality patterns.

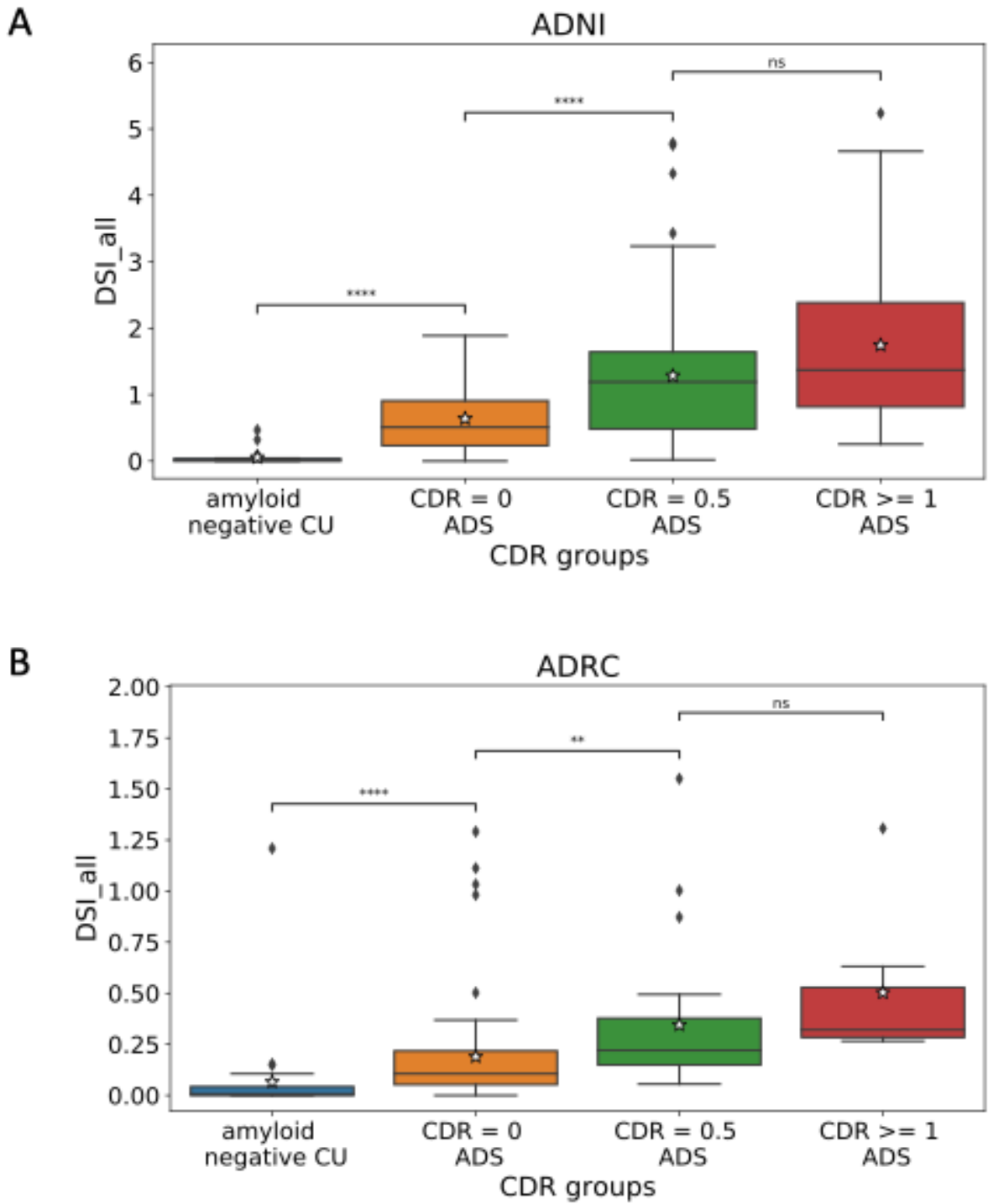

**Figure 5**: Box plot showing DSI_all (DSI across all modalities; see Section 2.5.5) for both ADNI (**5A**) and Knight ADRC (**5B**). The x-axis shows the different CDR groups in the ADS and CU-test (Section 2.2.1 and 2.3.1). FDR-corrected post hoc Tukey comparisons used to assess pairwise group differences. Abbreviations: DSI: Disease Severity Index, CDR = Clinical Dementia Rating. Statistical annotations: ns: not significant $0.05 < p <= 1$, * $0.01 < p <= 0.05$, ** $0.001 < p < 0.01$, *** $p < 0.001$.

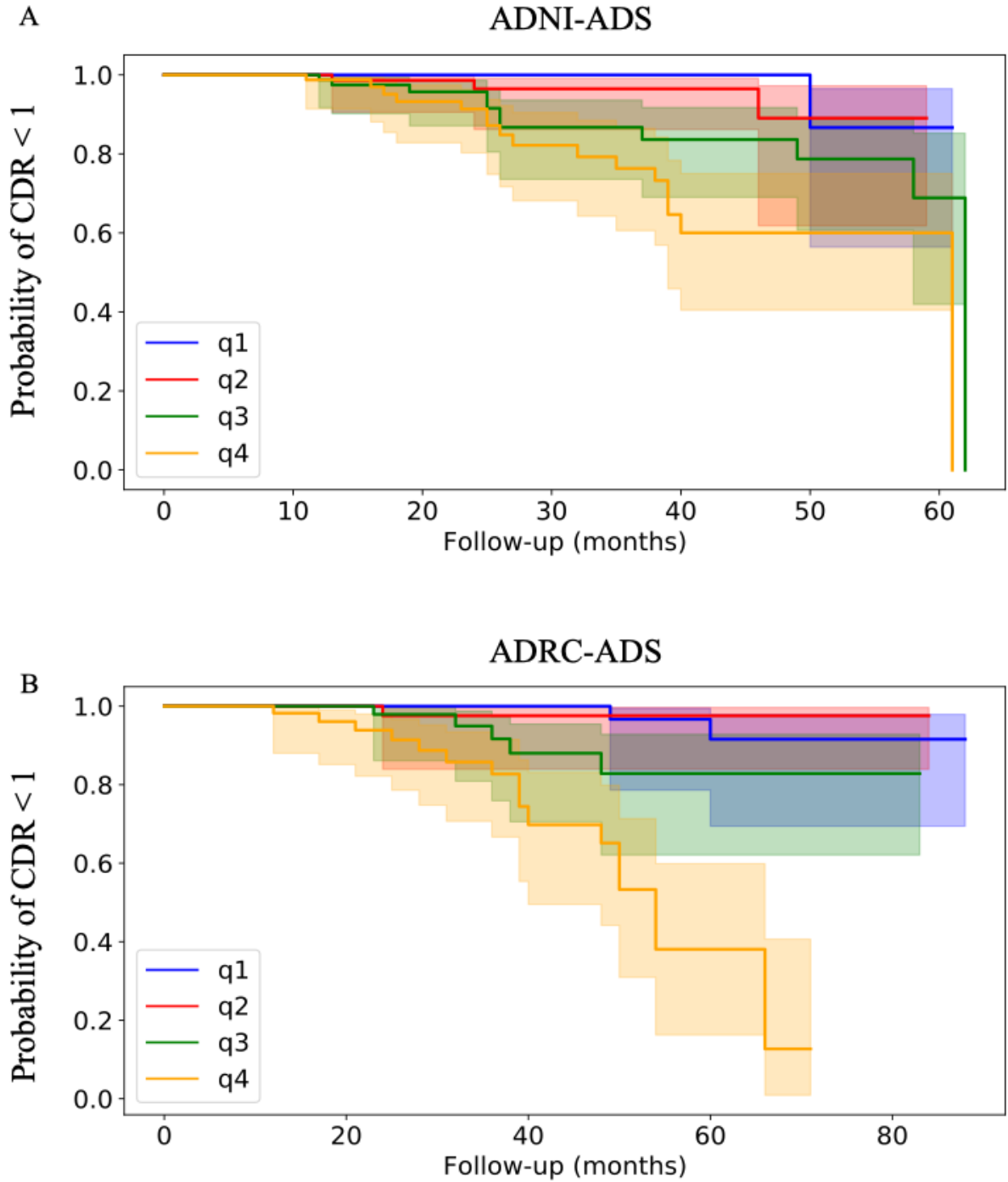

**Figure 6**: Kaplan-Meier plot of conversion from CDR < 1 to CDR >=1 for ADNI-ADS **(6A)** and ADRC-ADS **(6B)** participants. The x-axis and the y-axis represent the follow-up period (in months) and the probability of progressing from CDR <1 to CDR >= 1 respectively. The four lines represent the four quantiles of DSI_all (DSI across all modalities), shown by blue, red, green and orange respectively. The filled color span represents the 95% confidence intervals.